# Post-sampling crowdsourced data to allow reliable statistical inference: the case of food price indices in Nigeria

Giuseppe Arbia[1], Gloria Solano-Hermosilla[2], Fabio Micale[2], Vincenzo Nardelli[3] and Giampiero Genovese[2]

**Abstract:** Sound policy and decision making in developing countries is often limited by the lack of timely and reliable data. Crowdsourced data may provide a valuable alternative for data collection and analysis, e. g. in remote and insecure areas or of poor accessibility where traditional methods are difficult or costly. However, crowdsourced data are not directly usable to draw sound statistical inference. Indeed, its use involves statistical problems because data do not obey any formal sampling design and may also suffer from various non-sampling errors. To overcome this, we propose the use of a special form of post-stratification with which crowdsourced data are reweighted prior their use in an inferential context. An example in Nigeria illustrates the applicability of the method.

**Key words:** Post-stratification. Spatial sampling plans. Efficiency. Crowdsourcing.

## 1 Introduction

Timely and reliable food price data are essential for sound policy and decision making. In the last decade food crises in developing countries have made the need for up-to-date food prices more pressing. Up-to-date prices are essential for the measurement of poverty and vulnerability and to inform early warning systems of food insecurity. Food price data from throughout Africa are indeed available, but they are often characterized by a lack of continuity, insufficient coverage, delays in or absence of dissemination. Innovative data collection systems emerge as an economic statistics big challenge for today's organizations and the use of crowdsourced data may provide a valuable alternative for data collection. However, the statistical methods may not be adequate in the case of pure crowdsourcing where data collectors are volunteer, anonymous and of variable and unknown reliability. In this case, data derive from non-probabilistic, convenience samples where the anonymity of the participants might encourage cheating to maximize income so that reliability, accuracy and representativeness becomes of major concern. In this paper we propose a methodology to validate crowdsourced data so as to be able to produce timely and reliable food price indices.

[1]Universita' Cattolica del Sacro Cuore, Milan (IT) [2]
[2] Joint Research Center, European Commission, Directorate D, Sustainable Resources, Unit D.4 Economics of Agricult
[3] Università Bicocca, Milan (IT)

## 2   Quality methods for crowdsourced food prices

In this section we introduce a set of procedures to produce quality estimations based on crowdsourced data in social surveys. In general the objective of data validation is to ensure a certain level of quality of the final output of a data production process. The validation procedures developed here concern the accuracy and reliability, and the comparability and coherence components of the quality framework suggested by the ESS (ESS, 2015). Indeed, the accuracy of a crowdsourcing exercise may be undermined by both sampling and non-sampling errors. Sampling errors are connatural to crowdsourced data. In fact, lacking a proper sample design, the probabilities of inclusion cannot be calculated thus making inference problematic. In addition, non-sampling errors may also arise due to, e. g., possible measurement errors due to wrong interpretations of the collectors, time invariance of the prices (due e. g. to the repetition of the same value several times to avoid checking), self-selection of collectors, locational errors due to mistakes in recording the coordinates, endogeneity of the errors possibly induced by a correlation with other variables, non-independence of collectors (in that collectors, that may form a "cartel" reporting values negotiated among themselves and not observed regularly) and possible fraudulent activities (for example those submitting duplicate reports using multiple accounts). The proposed quality methodology involves two phases: a pre-processing phase, aiming at reducing non-sampling errors, and a post-sampling phase aiming at reducing sampling errors. In what follows we will discuss these two phases.

The pre-processing involves the identification and removal of standard and of spatial outliers. Spatial outliers differ substantially from standard outliers in that they represents a value that departs dramatically from the values observed in the neighbourhood. The identification of spatial outliers may help in spotting possible measurement errors due to wrong interpretations of the data collectors and possibly their non-independence and fraudulent activities. Let us call N(i) the set of neighbours of location i. Neighbours may be determined based on the pairwise distances between points which in turn can be acquired through the process of geocoding automatically obtained through Google Maps. Travel distances as well as travelling time on recommended routes can also be obtained through Google Maps Distance Matrix API. On the basis of this information, a weight matrix W is derived with a generic element given by 1 if $w_{ij} \in N(i)$ and 0 otherwise. From this definition, we also derive the spatially lagged value of a given point as the average values observed in its neighbourhood, which is further defined as $\text{lag}(P_i) = \sum_{i=1}^{n} w_{ii} P_i / \#N(j)$ with #N(j) the cardinality of the set N(j) and $P_i$ the variable (in our case price) observed at location i. A spatial outlier is finally defined as the value which exceeds r times the standard deviation the average values in its neighbourhood. After identification, the spatial outliers can be replaced with the average of the neighbouring observations.

The strategy employed to tackle the lack of a precise sampling design in crowdsourced data can be presented as follows. Suppose that *N* observations are crowdsourced in a given set of geographical locations. We can compare the map of the crowdsourced data with the map of points selected following a formal sample design of equivalent sample size. In each of the subareas considered, the *N* crowdsourced observations can be then reweighted so as to resemble the formal sampling scheme. We consider, in particular, two benchmarking designs: the case of a geographical stratification of the units and the case of an optimal spatial sampling design. As it is known, spatial sampling is distinct from conventional sampling due to the likely presence of spatial correlation amongst empirical data (Arbia, 2014). A popular strategy, initiated by Arbia (1993) consists in exploiting spatial correlation to maximize the information content while minimizing the sample size thus reducing the overall costs. Starting from this idea, a number of methodologies have been proposed that are based on an updating rule of the probabilities of inclusion. In particular, in this paper we consider the Local Pivotal method 2 (LPM2) (see Grafström et al., 2012). Having these two benchmarking sampling designs in mind, the proposed procedure develops through the following steps.

STEP 1: We start considering the observations of the variable of interest P in location $l$, say $P_l$. We assume that there are L locations in the study area, and that we have available, through crowdsourcing, $n_1, n_2 \ldots n_L$ observations in the L locations. The total number of observations in the whole crowdsourced exercise is N $= \sum_{l=1}^{L} n_l$ .

STEP 2: A formal sampling design (e.g. random stratified or LPM2) is then defined to select the same number (N) of locations as those observed through crowdsourcing. We call $m_l$ the number of locations thus selected in each location $l$.

STEP 3: We calculate, in each location, the *post-sampling ratio* (PS), defined as the ratio between the number of observations available through crowdsourcing and the number of observations required by the reference sampling design in each area. So, in each area l, we have $PS_l = m_l/n_l$.

STEP 4: The estimation of the mean of the variable of interest for the whole study area is then obtained as a weighted average of the observations in each location using the post-sampling ratio as weights.

Thus, if $PS_l$ = 1 then the number of observations available in location l are exactly those required by the sampling plan and no adjustment is needed. Conversely, if $PS_l$ > *1* then the number of observations available in location i are less than those required by the sampling plan and the observations available will be over weighted. Finally, if $PS_l$ < *1* then the number of observations available in location l are more than those required by the sampling plan and the observation will be underweighted in the average process. If no observations are available in location l ($n_l$ = 0), then the location is clearly not considered in the averaging process. Similarly, if no observations are required in location l by the sampling plan ($m_l$ = 0), then the observations collected in location l will also not contribute to the calculation of the global mean.

## 3   Simulation results

To test the performances of the suggested method we run a simulation study where we considered 1,000 individual points divided into 4 geographical strata characterized by unequal densities represented in the 4 quadrants of a unitary square (-0.5,0.5). The population size in the four strata is respectively given by 800, 60, 60, 80 (see Figure 1). Individuals' locations are randomly generated according the complete spatial randomness scheme (CSR, Diggle, 1983) in the 4 quadrants (see Figure 1). The variable of interest Y is generated using a pure spatial autoregressive model (Arbia, 2014) with spatial parameter $\lambda = 0.7$.

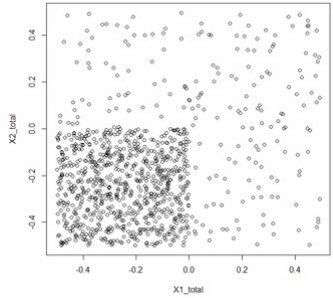

**Figure 1**: Location of 1,000 simulated observations in four quadrants of a unitary square.

In our simulation, we consider a sample of 80 individuals selected from the 1,000 population points. It is obviously difficult to simulate data that are generated without any regularity as it should be in a crowdsourcing exercise. In this paper we mimic a crowdsourced behaviour by considering a simple random sample of 20 units in each quadrant and we compare the performances of three distinct strategy. The first assumes the sample as a simple random sample and estimates the mean of variable Y with the Horvitz-Thompson estimator. This choice will obviously lead to unbiased, although highly inefficient estimators because it neglects the different densities with which data are distributed in the four quadrants. The second strategy uses the Horvitz-Thompson estimator with the individual data that are weighted using post-sampling ratios obtained comparing the actual data with a random stratified design with probability proportional to size. Finally the third strategy uses again the Horvitz-Thompson estimator, but with data that are weighted using post-sampling ratios obtained comparing the actual data with a LPM2 design. After 1,000 replications all strategies display very small absolute relative biases (with an average of 0.005). However, in terms of efficiency, the two post-sampling procedures largely outperform the simple random case (which, as said, mimic the crowdsourced data) producing a variance of the estimator that is 125 and, respectively, 124 times smaller than the one associated to the simple random case. The main results of the 1,000 replications are displayed in Figure 2.

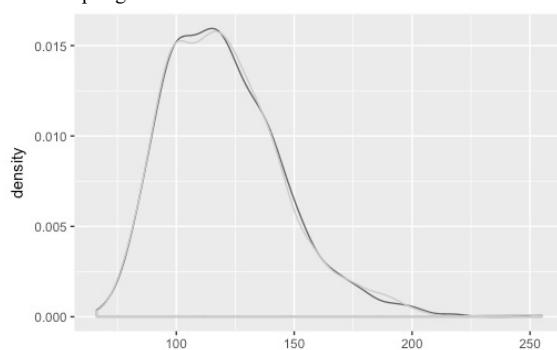

*Figure 2: MC distribution of the relative efficiency of the two post sampling strategies with respect to the simple random case in 1,000 replications. Darker shade = post sampling using stratified random sample as a reference, Lighter shade = post sampling using LPM2 as a reference.*

## 4 Data Analysis

In this section we report an application of our methodology to a set of Guinea Corn price data deriving from a crowdsourcing exercise run by AMIS-FAO in the period November 2016-March 2017 in 16 local markets of Kaduna State in Nigeria (Seid & Fonteneau, 2017). When missing, data were simulated using generated from a normal distribution with the same mean and standard deviation of the observed data. The first phase of our analysis involved the pre-processing: standard outliers were replaced with simulated data (only for the sake of our exercise), while spatial outliers were replaced by the average of prices in the neighbourhood. For the post-sampling exercise, we considered the 16 locations arranged into four spatial clusters which were identified through the k-means clustering algorithm (Forgy, 1965). Figure 1 shows the geographical position of the 16 locations considered and the four clusters that were created (identified by different colours). The large unboxed numbers in each cluster refer to the cluster id, while the small numbers into boxes (either circled or squared), refer to the location id. Locations identified with a circled number are those that were selected by the LPM2 method. In contrast, locations identified with a squared number represent those that were excluded from the optimal sampling design. Finally, locations identified with a crossed number are those for which we avail crowdsourced data. We then calculated the average price of Guinea Corn in Kaduna state as the weighted average of the clusters' means using the post-sampling ratio as weights. After post-sampling correction, the average price was 221.01 with an increase of 4.4 % with respect to the uncorrected average price (211.61).

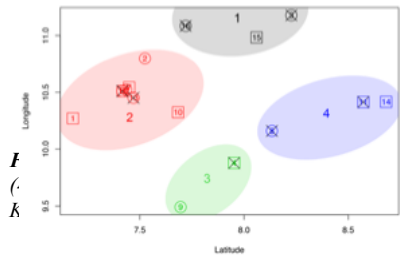

Figure ... na state (Nigeria). Codes: (1) Chikun, (2) Igabi, (3) Ikara, (...) (7) Kaduna north, (8) Kaduna state, (9) Kagarko, (10) ... Tasha, (14) Saminaka, (15) Soba and (16) Zaria

## 5 Conclusions

The present paper proposes a methodology consisting on a series of validation algorithms to produce quality food price indices. We tested the proposed method on real and simulated data showing its relative advantages. From a practical perspective the proposed quality methodology can be useful for institutions or organizations that aim at complementing the price data collection systems with crowdsourcing approaches. This methodology can be expanded and adapted to meet the needs of quality assessment of crowdsourced data in other regions and sectors.

## 6   Citations and References